\documentclass[10pt]{article}
\usepackage[papersize={8.5in,11in}]{geometry}
\usepackage{amsmath,amssymb,graphicx,paralist}

\newcommand{\comment}[1]{}

\begin{document}

\title{Comment on ``The Real Problem with MOND'' by Scott Dodelson, arXiv:1112.1320}

\author{J. W. Moffat$^{1,2}$ and V. T. Toth$^1$\\
{\rm\footnotesize $^1$Perimeter Institute for Theoretical Physics, Waterloo, Ontario N2L 2Y5, Canada}\\
{\rm\footnotesize $^2$Department of Physics, University of Waterloo, Waterloo, Ontario N2L 3G1, Canada}}

\maketitle

\begin{abstract}
We comment on arXiv:1112.1320 and point out that baryonic oscillations of the matter power spectrum, while predicted by theories that do not incorporate collisionless cold dark matter, are strongly suppressed by the statistical window function that is used to process finite-sized galaxy samples. We assert that with present-day data sets, the slope of the matter power spectrum is a much stronger indicator of a theory's validity. We also argue that MOND should not be used as a strawman theory as it is not in general representative of modified gravity theories; some theories, notably our scalar-vector-tensor MOdified Gravity (MOG), offer much more successful predictions of cosmological observations.
\end{abstract}


In \cite{Dodelson2011}, the author presents a case against MOdified Newtonian Dynamics (MOND) \cite{Milgrom1984} and, by implication, against other modified gravity theories, by asserting that these theories, MOND in particular, cannot possibly reproduce the power spectrum of matter, as derived from galaxy-galaxy correlations obtained from large scale galaxy surveys such as the SDSS \cite{SDSS2004, SDSS2006}.

While it is indeed true that the matter power spectrum might be used in principle to distinguish modified gravity theories from dark matter models, as we pointed out first in 2007 \cite{Moffat2007c,Moffat2007e}, the situation is not as simple as argued in \cite{Dodelson2011}.

Indeed, in the absence of dark matter, the unit baryonic oscillations in the matter power spectrum are not dampened by any physical process. They should indeed be visible in the matter power spectrum, {\em provided an infinitely large data set is used}. In reality the data set, while large (of order $10^6$ galaxies), is nonetheless finite, and the galaxy-galaxy correlation function is estimated by applying a suitably chosen window function to the data.

A realistic comparison of the matter power spectrum against the predictions of a theory cannot be obtained, therefore, unless the predicted power spectrum itself is convoluted with the same window function to simulate its effects.

\begin{figure}
\begin{center}
\includegraphics[width=0.75\linewidth]{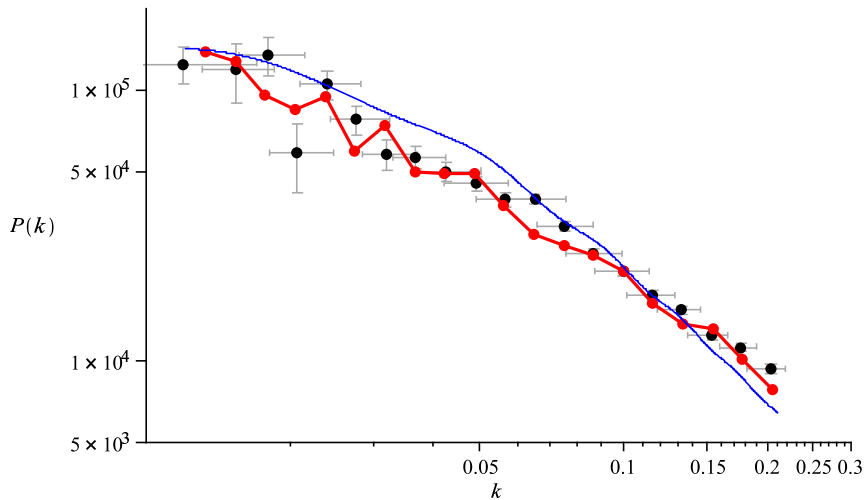}
\end{center}
\caption{The effect of window functions on the power spectrum is demonstrated by applying the SDSS luminous red galaxy survey window functions to the MOG prediction. Baryonic oscillations are greatly dampened in the resulting curve (solid red line). A normalized linear $\Lambda$CDM estimate is also shown (thin blue line) for comparison. From \cite{Moffat2007c,Moffat2007e}}
\label{fig:mm}
\end{figure}

We applied such a convolution to the predictions of our scalar-tensor-vector theory of MOdified Gravity (MOG) \cite{Moffat2006a} and obtained the curve shown in Fig.~\ref{fig:mm}. This figure clearly demonstrates that when realistic sample sizes are taken into account, convolution of the window function with the theoretical prediction effectively removes most baryonic fluctuations, making it impossible to exclude theories on this basis alone.

What was not discussed in \cite{Dodelson2011}, however, is the fact that in addition to baryonic fluctuations, another characteristic that can strongly distinguish modified gravity theories from dark matter models is the slope of the matter power spectrum curve. Even if baryonic oscillations are removed by convolution with the window function, a theoretical prediction may have a much different slope than the observed matter power spectrum. This is the case when the matter power spectrum is predicted using Einstein gravity and baryonic matter alone, and, as evident from Fig.~1 in \cite{Dodelson2011}, this is indeed also the case for MOND.

MOG, however, fares better, as the theory includes a variable gravitational constant that manifests itself as an additional power law dependence in the matter power spectrum, tilting the curve in the right direction. We argue that given the availably galaxy data sets at present and in the immediate future, the slope of the matter power spectrum is a much stronger predictor of the likely success or failure of a gravity theory than baryonic oscillations, as the latter are suppressed by the window function associated with the finite size of the data set.

Lastly, we would like to comment on the unfortunate trend in the community to use MOND as a strawman for modified gravity theories in general. MOND is not a theory: it is a phenomenological formula that fits some data (notably, galaxy rotation curves) and fails elsewhere. Attempts to create a proper gravitational theory that yields MOND in the weak field limit, such as TeVeS \cite{Bekenstein2004}, are, however clever, nonetheless contrived. The failure of these attempts to account for a wide range of cosmological observations must not be taken as indication that a properly constructed, physically motivated gravity theory, such as MOG, cannot live up to the challenge presented by modern-day precision cosmological and astrophysical observations at least as well as the prevailing $\Lambda$CDM concordance cosmology.

\bibliography{refs}
\bibliographystyle{unsrt}

\end{document}